# Turbulent shear flow of solids

Y.Beygelzimer[1], A.E.Filippov[2,3,4], R.Kulagin[5,*], Y.Estrin[6,7,*]

[1]Donetsk Institute for Physics and Engineering named after A.A. Galkin, National Academy of Sciences of Ukraine, Nauki ave., 46, 03028 Kyiv, Ukraine.
[2]Functional Morphology and Biomechanics, Zoological Institute, Kiel University, Am Botanischen Garten, 1-9, Kiel 24118, Germany.
[3]Donetsk Institute for Physics and Engineering, NASU, Donetsk, Ukraine.
[4]Technische Universität Berlin Institut für Mechanik FG Systemdynamik und Reibungsphysik Straße des 17. Juni 135, 10623 Berlin, Germany.
[5]Institute of Nanotechnology, Karlsruhe Institute of Technology, Hermann-von-Helmholtz-Platz 1, 76344 Eggenstein-Leopoldshafen, Germany.
[6]Department of Materials Science and Engineering, Monash University, 22 Alliance Lane, Clayton 3800, Australia.
[7]Department of Mechanical Engineering, The University of Western Australia, 35 Stirling Highway, Perth WA 6009 Australia.
[*]Corresponding Authors' E-Mails: roman.kulagin@kit.edu, yuri.estrin@monash.edu

**Statement of Significance**

The notion of 'solid state turbulence' may go against common sense, but the phenomenon does exist! We demonstrate that the very essence of solids – their ordered structure enabling stability of shape – may lead to the occurrence of turbulent shear flow of solids.

**Abstract**

The term 'solid-state turbulence' may sound like an oxymoron, but in fact it is not. In this article we demonstrate that solid-state turbulence may emerge owing to a defining property of the solid state: the ability of a solid to retain its shape. We consider shear flow of layers of solids with different stiffness and show that the stiffer ones may spontaneously decompose into a set of blocks. This breakdown of isometry is a key to plasticity of solids and is fundamental for the occurrence of solid-state turbulence. To visualise the piecewise isometric transformations of the blocky structure in a turbulent flow regime, we use a heuristic model based on discretisation of a continuum into interacting 'particles'. The outcomes of the numerical experiments conducted support the occurrence of pulsations of velocity and pressure in plastically deforming solids and the emergence of vortices characteristic of classical turbulence. This phenomenon may have important practical implications for solid-state mixing as an ecologically beneficial alternative to conventional metallurgical processing routes.

**Introduction**

A flow which exhibits irregularities, chaotic pulsations of velocity and pressure, and random vortices of different scale that lead to mixing of the constituent substances is referred to as turbulent [1]. Almost any turbulent flow of a fluid is of shear character, its layers moving past each other under a shear force. Therefore, turbulence in shear flows is the main problem considered in that context [2]. Of special interest is the case of free shear flow, i.e., the flow of a fluid confined within another fluid.

It is almost a platitude to say that turbulence is one of the greatest enigmas of Nature, but this is, indeed, an apt statement [3]. Commonly, turbulence is associated with the gaseous or liquid aggregate states only. However, in metals deformed by severe plastic deformation (SPD) [4] a phenomenon clearly showing features of turbulence was observed. Similar effects are found in near-surface layers of metals engaged in dry friction [5, 6]. In what follows, we shall refer to this phenomenon as *solid-state turbulence* (SST).

We would like to mention that as early as in 1953, Cottrell used the term 'turbulence' to describe a transition from the 'laminar' dislocation motion on a single slip system in stage I of strain hardening of metals to multiple slip when several slip systems get activated in stage II [7]. Unlike this usage of the term as a figure of speech, we shall be talking about turbulence in a genuine sense, as a phenomenon characterised by chaotic changes in the flow velocity and the occurrence of unsteady vortices.

The article [8], in which a vortex structure in titanium processed by twist extrusion was documented, appears to be a first reported case of SST induced by SPD. The occurrence of the vortices was associated with rotational motion of the material under simple shear at the boundaries of the screw-shaped part of the twist-extrusion die. The author of [9] invoked an analogy of SPD-induced SST with turbulence in fluids and hypothesised that vortices in a deforming solid are generated by obstacles hindering shear flow. Recently, several articles dealing with SST were published, but literature on the subject is still scarce. Thus, it was proposed to employ SST for mixing of metals aimed at fabricating new alloys or architectured materials. The principal SPD technique used to that end was high-pressure torsion (HPT) [10]. The process involves shear flows of two or more metals giving rise to vortices at their interfaces, which lead to random oscillations of the flow velocity and vigorous mixing of the metals, cf., e.g., [11-13]. The flow pattern and the shape of inclusions of the harder of the constituent metals within the relatively soft matrix resemble the picture observed in free shear flows of fluids. This was the motivation for computational modelling of the emergence of turbulence in a shear flow of multilayered metallic laminates in terms of the rheology of nonlinear viscous fluids [14] by finite element analysis [15, 16] and molecular dynamics simulations [17, 18]. Establishing a quantitative measure of the efficacy of mixing under SST, proposed and substantiated in [19], is also to be mentioned in this context. Although the possibility of the occurrence of solid-state turbulence may be evident from these reports, it is fair to say that the scientific community is largely unaware of the existence of the phenomenon. At any rate, the mechanisms of SST are not understood, and it appears timely to address them.

Investigations of SST as a stochastic nonlinear phenomenon is of great scientific significance, as it has certain universality and can occur at different length scales: from laboratory experiments to processes in the Earth's lithosphere [20]. The phenomenon is also of practical importance, as SST opens an avenue for solid-state mixing and physicochemical reactions under high pressure, whose potentialities in materials engineering can hardly be overestimated [13, 21]. Indeed, SST offers an ecologically advantageous alternative to conventional metallurgical processing.

In this article, we propose and corroborate a hypothesis on the nature of SST and investigate some salient features of this phenomenon. To that end, a computational heuristic model was developed and applied. Fermi, Pasta, and Ulam appear to be the first to use a computer as a heuristic tool for solving a complex physical problem [22]. In the work that followed, Ulam and Pasta broadened the applications of this approach. Nowadays heuristic computer models are widely used to solve complex problems in a variety of research fields [23, 24].

Working with a heuristic model is like playing an exciting game in which the researcher can observe the behaviour of a complex system in different situations created at will. This makes it possible to hypothesise on the origin of the phenomenon in question, identify the major governing parameters and their interplay, and get insights in the underlying physics, which enables development of quantitative models. For the 'game' to be efficient in this regard, the heuristic model must account for the salient features of the phenomenon meaning that it must capture the main governing factors. Accordingly, we start by establishing the factors that determine SST.

1. **The major factors defining SST: modelling preliminaries**

Since SST is found for solids of any chemical composition and size, we consider the principal defining characteristic of the solid state – the ability of a solid to retain its shape. Let us express this property in a formal way, so that it can be implemented in a model.

To be specific, we first look at crystalline solids. Their atomic lattice, which is responsible for long-range order in the system, is what distinguishes them from liquids. The existence of the crystal lattice owes to the physics of the interatomic interactions. We take this as a departure point of our analysis and pose the problem of deformation of a solid in geometrical terms. This will allow us to draw some rather general conclusions regardless of the chemical or phase composition of the solid.

We denote the initial and the final configurations of a solid body by A and B, respectively. The geometrical operation which transforms A to B is denoted by *G* [25]. It is known that plastic deformation has no substantial effect on the lattice parameter, changing it by no more than ~0.1%. This means that if *G* is to represent plastic deformation, it can be considered to belong to the class of *isometric transformations*. These are transformations under which the distance between any two points in the body remains unchanged [26]. Transformations of this kind include rigid displacements, rotations, and mirror reflections, see Fig. 1.

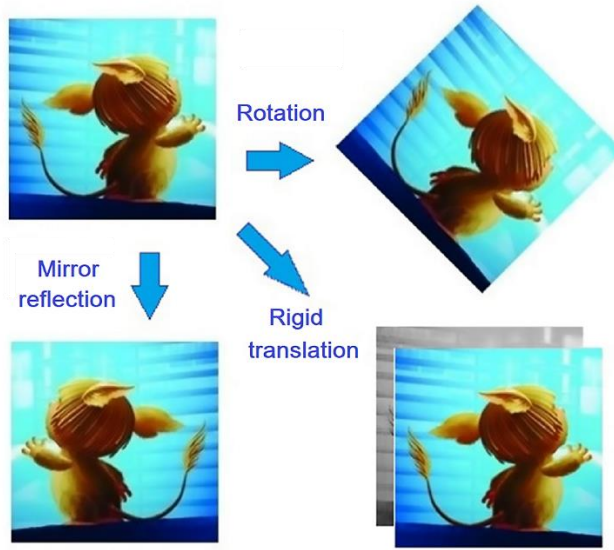

**Figure 1.** Examples of isometric transformations.

If the solid does not undergo plastic deformation, the transformation *G* is the same for its entire volume, so that the configurations A and B differ only in the spatial location and/or orientation. In the opposite case, *G* must include discontinuities of displacements, rotations, or mirror reflections, which puts it in the category of isometric transformations with singularities, commonly referred to as *piecewise isometries* [27].

So far, we considered only crystalline materials, but the conclusion that the geometry of a solid undergoing plastic deformation changes by means of piecewise isometries has general validity and applies to amorphous solids, as well. It also applies to fracture processes, which occur by subdivision of a solid into fragments that are isometric to their original regions within the solid before fracture. Comminution of a rock is an example of such a process.

The concept of piecewise isometry has been applied to describe various phenomena and systems, such as, e.g., pattern formation in Nature [28], mixing of powders [29-31], and behaviour of dynamic systems [27].

Discontinuities of isometry during deformation and fracture of solids can occur at different length scales. At the atomic scale discontinuities in the displacements, rotations, and mirror reflections are furnished by dislocations, disclinations, and twins, respectively. At meso and macro scale, isometry discontinuities can manifest themselves in stationary or propagating shear bands, including Lüders bands, and other strain localisations accompanying plastic deformation [32-36]. At a larger length scale, cracks are a signature of isometry discontinuities leading up to fracture.

The concept of plastic deformation and fracture of solids being describable in terms of piecewise isometries provides explanations for a whole range of observed effects. As an example, isometry discontinuities associated with the deformation of solids have a stochastic character and can give rise to mass transport and efficient mixing of co-deforming solids. We believe that isometric discontinuities may also be one of the possible reasons for anomalously fast diffusion in severely deformed metals [37].

By the same token, we see the origin of SST in piecewise isometries accompanying deformation and ruptures of a solid. The maximum distance between singularities of isometry determines the spatial scale of SST. The characteristic size of the isometric fragments of a solid is governed by the constraints put on the flux of the material. For example, if a solid is sheared between two rigid surfaces, it will imitate laminar flow, the deviations from laminarity becoming more and more evident as the length scale of the observation is decreased. If the constraining boundaries are not rigid, the non-laminar flow will disturb them more and more as the amount of shear increases, so that turbulence will become manifest at macro scale.

We believe that it is a tight constraint on deformation in HPT that explains why it results in the smallest grain size (i.e., the greatest degree of fragmentation) among the SPD techniques [4].

Thus, the first and foremost tenet of a prospective heuristic model of solid-state turbulence is the assertion that the flow geometry of a solid body is associated with piecewise isometries. A second important ingredient of the model is the establishment of conditions leading to shape variation of a solid, i.e., to the generation and propagation of isometry discontinuities. In solid mechanics, such conditions correspond to plasticity and fracture criteria. A rather general criterion of this kind is that of a critical magnitude of the elastic strain energy, which is related to the yield stress under shear deformation.

In the next section, we present our mathematical model of solid-state turbulence. The model may be somewhat naïve compared to traditional models of plasticity and fracture, yet it will be shown to have the advantage of accounting for the salient features of SST.

## 2. The model of solid flow

Our aim was to employ a model of plastic flow in solids that would be robust and simple, yet capable of capturing the above principal factors defining SST. Based on such a model, we endeavoured to demonstrate through numerical experiments that it is the patchiness of the emerging geometrical patterns that breaks the laminarity thus giving rise to turbulent flow. The patchiness, in turn, is contingent on the degree of order in the system. We are obviously dealing here with a particular case of a general problem: how do meso- and macroscopic properties (including the elastic and plastic ones) and the associated geometrical patterns derive from dynamic processes described at a deeper theoretical level? Suitable modelling tools are provided by discretisation of a continuum and tracing the dynamics of 'particles' representing it. For example, the model proposed in [38] in terms of the dynamics of discrete 'particles' yields a broad range of mass and energy transport regimes – from the solid to the gaseous state - for the simplest case of a repulsive interaction potential between the 'particles'. This approach provides control of the correlation radius between the elements of an effective continuum, i.e., the degree of order, which is crucial for the occurrence of SST. To avoid confusion, it should be stated from the outset that the use of the notion of 'particles' here and below does not imply that individual atoms are considered. Recently, a significant progress with understanding structural lubricity in soft and hard matter systems was achieved by using a similar modelling approach [39]. At present, it is still not possible to advance to the satisfactory length and time scales by using classical molecular dynamics that operates with atoms or molecules. That is why modelling approaches in which representative larger-scale objects whose motion in its entirety reflects the pertinent aspects of the behaviour of the system have come to the fore. These include, for example, the method of moveable cellular automata [40], smoothed particle hydrodynamics [41], and the discrete automata method [42, 43]. Not only do these approaches show a qualitative accord between modelling and experiment, but they also enable a good quantitative agreement for various systems considered. However, they are computationally costly. Besides, visualisation and processing of the results of the computations are very involved [44].

Our concern in the present work was primarily the extraction of qualitative information from the numerical simulations, with a convenient access to direct visualisation for various scenarios and over large periods of time. As a method of choice, we used the isotropic 'moveable automata' technique (cf., e.g. [45]), which involves a weak long-range attractive interaction between particles in addition to a short-range repulsive one. The form of the interparticle interactions used ensures the existence of a minimum of the overall potential and the attendant equilibrium state. Such a system does not require 'walls' created by the boundary conditions to form compact fragments of a crystalline lattice. Even for the case when the average density of the material is lower than that required for contiguous space filling, the particles show a tendency to aggregation. The freedom of the solid to deform plastically is still retained, as the lattice is subdivided in fragments capable of moving over distances far greater than the lattice constant. In doing so they can rotate and deform elastically nearly as a whole, or fuse with other fragments that may have been distant initially. The ability to rupture enables the lattice to engulf particles of other substances that were separated from it. Naturally, the combinations of attraction and repulsion are different for the sub-systems of a two-component or, generally, multicomponent system, as well as between the subsystems. It can be shown that a strong mutual repulsion promotes separation of subsystems, whereas an attraction enhances the tendency to mixing. We used this observation in connection with modelling of the behaviour of materials with different stiffness that show some (but not exceedingly strong) proclivity for separation. The details of the model used are given in the Supplementary Information.

In solving the equations of motion, Supplementary Information Eq. (S3), we confined ourselves to problems where the thermal energy $k_B T$, where $T$ is the absolute temperature and $k_B$ is the Boltzmann constant, is negligibly small compared to other energy terms involved. It can be proven [38], that in the limit of high kinetic energy density, $E_{kin} \gg C_{ij}$, where $C_{ij}$ is a stiffness parameter associated with the magnitude of the attractive interaction between particles $i$ and $j$, the system described by Supplementary Information Eqs. (S1) and (S2) behaves like a gas of nearly free-flying particles. In the opposite limit case, $E_{kin} << C_{ij}$, a crystal lattice is formed spontaneously. The lattice constant $a$ is determined by the equilibrium distance between the particles. The nodes of the lattice oscillate around the bottoms of the potential valleys formed by the neighbouring particles. This oscillatory motion can be described in terms of a harmonic Hamiltonian slightly perturbed by non-linear terms. In this sense, the system deforms elastically. While in the gaseous state the system is isotropic on average, in the solid state its lattice has a hexagonal symmetry if the interaction potential between the particles is isotropic. In both limit cases, the system exhibits a regular dynamic behaviour. However, due to a difference in symmetry, there is a transition between the two states via a mixed disordered state, which, in the presence of dissipation, represents a viscous fluid. Thus, the minimalist model considered accounts for the occurrence of all three aggregate states of the system.

3. **General features of solid-state flow as revealed by computational experiments**

Based on the model of solid-state flow outlined above and detailed in Supplementary Information 1, we conducted computational experiments and visualised their results. As the system consists of discrete moveable particles, the following computational procedure was used for our numerical experiments. A certain number of particles were placed on a rectangle with the sides $[0, L_x]$ and $[0, L_y]$ at random at near-zero temperature. Due to their interactions described in Section 2, the particles formed a near-equilibrium structure consisting of a set of differently oriented domains. The system was then engaged in shear flow by the horizontal movement of the upper and lower plates in the opposite directions. For laminar flow, the velocity distribution along the $y$-axis of particles trying to follow the plates would become a smooth monotonic function. This is not possible, however, if the solid is sub-divided into domains. It will first deform elastically, the lattice nodes following the displacement of the plates smoothly. This is reflected in a smooth distribution of the horizontal velocity over the ordinate axis, $v_x(y)$. Once a critical shear strain is exceeded, the bonds between some of the nearest neighbours start breaking. Other particles now become neighbours. Although formally all particles of the system interact with each other, in reality a particle is enslaved by its immediate neighbourhood and tends to follow it. This means that locally a fragment of the lattice moves as a whole swapping crystallographic planes only at contacts between domains. This represents the characteristic behaviour of solids supporting large strains by means of piecewise isometric transformations. The results confirm the validity of the two cardinal principles which, as discussed in Section 2, determine the essence of SST: (i) plastic deformation is mediated by piecewise isometries and (ii) isometry discontinuities set in once a certain critical magnitude of the shear force (corresponding to a critical elastic strain energy) is reached.

A typical picture of the particle movement of the kind described revealed by computer experiments is visualised in the video clip Supplementary Movie_01.avi. The red and blue circles in the spatial distribution map correspond to positive and negative projections of the velocity on the abscissa, respectively. Contrast spots and bands give a clear picture of instantaneous velocity bursts in proximity to the lines separating the regions where movement of the particles occurs with nearly the same velocity. These are the lines where switching of nodes between crystallographic planes occurs. These lines can be interpreted as isometry discontinuities with the greatest level of energy dissipation.

In addition to the velocity distributions, it is also instructive to use the Voronoi and Delaunay diagrams. A Voronoi diagram for a set of points (particles) on a plane is a tessellation of the plane into regions where each point is closer to all points within that region than to any point of the set outside of it. A Delaunay diagram is a triangulation of the same set of points such that for any triangle all points, except those at its apexes, are situated outside of its circumcircle. The Voronoi diagram permits identification of the symmetry of the neighbourhood of any point, while Delaunay triangulation, which is dual to the Voronoi mesh, enables visual construction of a lattice at any step of a numerical experiment. Figure 2 displays a typical instantaneous snapshot of a system represented by the Voronoi and Delaunay constructions. One can readily recognise a subdivision of the system into domains and chains of defects with 5-fold and 7-fold symmetry axes (coloured blue and orange, respectively). A blow-up in the insert presents an enlarged view of the Delaunay lattice where some of the defects are highlighted by pentagons for clarity.

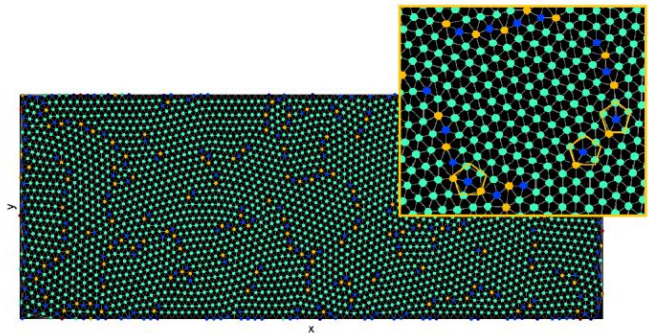

**Figure 2.** A typical snapshot of a 'crystalline' system of particles by means of the Voronoi and the Delaunay diagrams. The enlarged picture in the insert facilitates the visualisation of the constructions described in the text.

It should be noted that such geometrical constructions are helpful both for qualitative presentation and for quantitative evaluation of the numerical simulation results, e.g., in the form of uniquely determined histograms for particle distributions.

4. **Solid-state turbulence as revealed by computational experiments**

Building upon the knowledge of the basic properties of the model discussed above, we can now turn to considering turbulence in a composite system comprised by *materials with different stiffness*. We construct the initial system as a layer of a stiff material sandwiched between two layers of a more compliant material. The latter are set in motion by two rigid plates shifted horizontally in the opposite directions with the same constant velocity, $V = const$. Figure 3 shows the results of numerical simulations.

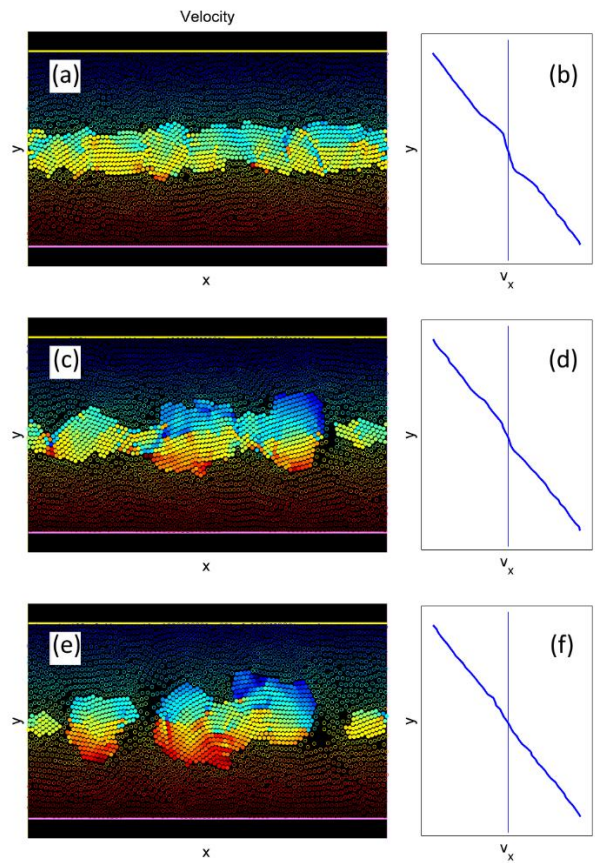

**Figure 3.** Turbulent flow of a layer of a stiff material sandwiched between two layers of a softer, more compliant material. The particles comprising the stiff and the pliant materials are represented by filled and open circles, respectively. The magnitude of the velocity along the horizontal axis is visualised in (a), (c), and (e) using the colours of the standard 'jet' colour code, in which red corresponds to the largest and blue to the smallest magnitude of a variable. The respective shear strains are roughly 0.5, 5, and 10. The diagrams (b), (d), and (f) show the distributions of the magnitude of the averaged velocity along the vertical axis, *y*.

As could be expected from preliminary numerical experiments, the outer layers transmit the motion into the bulk of the sandwich relatively quickly, the stiff inner layer attaining a velocity of a substantial magnitude. The corresponding variation of the velocity with the depth is smooth and almost linear in the pliable outer layers and abrupt in the hard inner layer, Fig. 3. An animation, Supplementary Movie_02.avi, clearly illustrates that at this initial stage the movement of the external plates is transmitted only to the more compliant sub-system, the stiff one being practically indifferent to this movement. Since the sandwich structure was created naturally, through spontaneous ordering of the layers of two different kinds before shear deformation, the interfaces between the layers exhibit some roughness. A phase separation between the layers is furnished by mutual repulsion between the particles of the two kinds considered to be the same as that between the particles of the same kind, combined with a smaller (or even non-existent) attraction. The smoothness of an interface and the separation distance between the layers of different kinds (or the opposite – the interdiffusion of the two particle species) are controlled by the magnitude of the attractive interaction.

To be specific, here we confine ourselves to an intermediate case when a tendency to phase separation does exist but is not excessive. Such a case is shown in Fig. 3c. Domains of different orientation are easily recognised. One can discern differently oriented domains within the stiff material and ordered (albeit more loosely) domains of the more compliant material lining the boundaries of the stiff layer without penetrating them.

A qualitative analysis of the deformation process at a later stage characterised by rupturing of the inner layer appears to be a good representation of the reality. The development of the process with time leads to deformation of the inner layer and the attendant smoothening of the distribution profiles of force and velocity integrated over the horizontal axis. Stress (or force) concentrations, as well as the associated waves of velocity, in proximity to the nascent discontinuities visualised in colour are readily recognisable. Details of the entire process can be observed in the video Supplementary Movie_02.avi. A late stage of the process after the fragmentation of the inner layer into separate islands is shown in Fig. 3e. Further smoothening of the extrema in the distributions of the force and especially velocity (where they disappeared altogether) is obvious. A characteristic velocity distribution within the islands indicates their practically independent solid-body rotation within the overall laminar flow.

Solid-state turbulence promotes mixing of the two materials of which the sandwich structure is composed. The degree of mixing can be quantified using a criterion proposed in [19]. It is based on the use of the mixing index $Q = [(G - G_{min})/(G_{max} - G_{min})]^{1/2}$ defined in terms of the so-called Gibbs parameter $G$, which in the present context is given by $G \equiv \sum_{k=1}^{N_{tot}} C_k^2 / N_{tot}$. Here $C_k$ is the distribution density of the stiff component over the nodes $k$ of a stationary grid and $N_{tot}$ is the total number of the nodes. $G_{min}$ and $G_{max}$ are the minimum and the maximum values of the Gibbs parameter, respectively. The minimum value, $G_{min} = C^2$, is attained for a uniform distribution, $C$ = const, while a maximum is reached when the concentration at each node of the grid is equal to zero or unity. Hence, the index $Q$ can be used to quantify the degree of mixing of the two components. We used this procedure to characterise the disruption of a stiff layer confined by two compliant layers described above. Figure 4 shows a snapshot of the system taken at an intermediate stage of its shear flow, along with the time dependence of the mixing index $Q$. It is seen that, as expected, the mixing index gradually drops with the progress of the shear flow process.

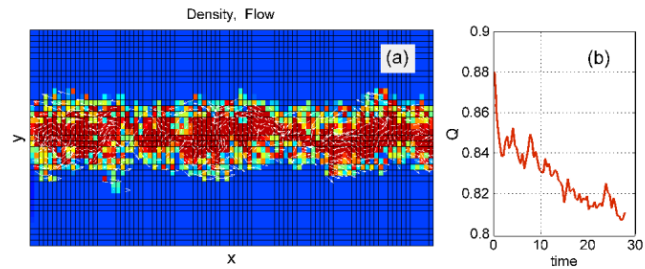

**Figure 4.** An instantaneous distribution of the density of the stiff component on a stationary spline grid (a) and the time dependence of the mixing index $Q$ (b). In addition, the vector velocity field reflecting the vortex-like behaviour of the flow is shown in (a).

**Discussion**

The results of the numerical experiments conducted to model SST allow us to draw some general conclusions and establish the salient properties of free shear turbulence in solids.

As shown in Sections 3 and 4, long-range order in the system gives rise to a blocky structure of both materials of a 'sandwich'. In the shear flow, the blocks adjust to each other, get fragmented or agglomerate, but most of their bulk undergoes isometric transformations, rather than deforming. Hence, the model underlying the numerical simulations corroborates our hypothesis of piecewise isometric character of the deformation of solids. The success of the model with adequately emulating SST gives reasons to believe that this characteristic feature of the deformation of solids is the root cause of solid-state turbulence.

What features of a real experiment correspond to the fragmentation of solid flow into blocks? It can be said with certainty that these blocks are not microstructural entities of the material (such as grains or sub-grains). Indeed, the solid considered in the model is structureless. The blocks represent chunks of material that are carried by the overall flow without being deformed. Their characteristic size, and the spatial pattern of SST in general, is dictated by the length scale of the order in the system and reflects the level of freedom of motion of the material points. Computer experiments presented in Section 3 showed that an increase of the block size is associated with the growth of the yield strength of the model material.

When the shear flow of a solid is constrained by rigid boundaries, segmentation of the blocks down to sizes much smaller than the width of the shear layer occurs. Viewed at macro scale, the solid-state flow appears to be laminar, but at smaller length scales disruption of the laminar flow becomes evident. If the shear flow of a solid layer occurs in a relatively compliant environment, the big blocks into which it gets subdivided are less prone to further fragmentation than in the case when the layer is confined by rigid walls. In the latter case, the turbulent character of the flow is manifested at a length scale prescribed by the thickness of the layer, and SST is pronounced at macro scale. This macroscopic turbulent flow leads to intensive mixing of the rigid and the compliant constituents of the sandwich. The deformation of a model material is caused by an external shear force, which is transferred into the bulk of the flux through interaction between particles. Since particles are agglomerated in blocks, an entire block acted upon by a force from its environment can be considered. According to the laws of mechanics, this force produces a central force and a momentum, which leads to block rotation. The model employed here accounts for this effect, which enables a qualitative description of the phenomenon of SST. To treat SST in a continuum mechanics model, gradient theories of plasticity would need to be invoked and a complex non-uniform stress-strain state of a solid under a shear flow would have to be considered [46].

## 5. Conclusion

In this article we hypothesised that the phenomenon of solid-state turbulence is a result of the fundamental property of the solid state: the ability of a solid to retain its shape and to change it only under sufficiently high stresses. To validate this hypothesis, we employed a discretised, multi-particle model of a shear flow of a solid, which spontaneously decomposes into a system of blocks by means of self-organisation of particles. During plastic flow, these blocks undergo piecewise isometric transformation, which is a core feature of deformation of a solid at large strain. The model is an oversimplification, in that it does not consider the chemical and phase composition of the solid. Neither does it reflect the real interaction of mobile particles in the solid. Despite its minimalism, or perhaps thanks to it, this model made it possible to mimic the behaviour of the system at the temporal and spatial scales not accessible to molecular dynamics simulations. The existence of turbulence in plastically deforming solids was demonstrated and several non-trivial results were obtained. The most important of them is the recognition that the main property of the solid state described above is sufficient for the occurrence of stochastic vortices in a solid-state flow. The effect of the type of the constraints put on the shear flow of a solid on the length scale of the turbulence patterns formed was also elucidated.

The model used in the numerical simulations is a heuristic one. It is simple and user-friendly and can readily be applied for future use aimed at testing new hypotheses relating to the nature of solid-state turbulence.

**Acknowledgements**

Support from Karlsruhe of Technology through the KIT Publication Fund is acknowledged.

**References**

1. Batchelor, G. K. *An introduction to fluid dynamics* (Cambridge Univ. Press, Cambridge, UK, 2000).

2. Smith, K., Caulfield, C., & Taylor, J. Turbulence in forced stratified shear flows. Journal of Fluid *Mechanics* **910**, A42 (2021).

3. Vallis, G. K. Turbulence theory: Imperfect, but necessary. *AGU Advances* **2**, e2021AV000523 (2021).

4. Valiev, R.Z. et al. Producing Bulk Ultrafine-Grained Materials by Severe Plastic Deformation: Ten Years Later. *JOM* **68**, 1216-1226 (2016).

5. Rigney, D.A., Karthikeyan, S. The Evolution of Tribomaterial During Sliding: A Brief Introduction. *Tribol. Lett*. **39**, 3–7 (2010).

6. Cihan, E., Störmer, H., Leiste, H., Stüber M. & Dienwiebel M. Low friction of metallic multilayers by formation of a shear-induced alloy. *Sci. Rep*. **9**, 9480 (2019).

7. Cottrell, A.H. *Theory of dislocations* (Oxford Univ. Press, Oxford, 1953).

8. Stolyarov, V., Beigelzimer, Y., Orlov, D., & Valiev, R.. Refinement of Microstructure and Mechanical Properties of Titanium Processed by Twist Extrusion and Subsequent Rolling, *The Physics of Metals and Metallography* **99**, 204-211 (2005).

9. Beygelzimer, Y. Vortices and Mixing in Metals during Severe Plastic Deformation. *Mater. Sci. Forum* **683**, 213-224 (2011).

10. Zhilyaev, A. P. & Langdon, T.G. Using high-pressure torsion for metal processing: Fundamentals and applications. *Prog. Mater. Sci*. **53**, 893-979 (2008).

11. Han, J. K., Herndon, T., Jang, J. I., Langdon T. G. & Kawasaki M. Synthesis of Hybrid Nanocrystalline Alloys by Mechanical Bonding through High-Pressure Torsion. *Adv. Eng. Mater*. **22**, 1901289 (2020).

12. Krämer, L., Champion, Y. & Pippan, R. From powders to bulk metallic glass composites, *Sci. Rep*. **7**, 6651 (2017).

13. Hernández-Escobar, D., Kawasaki, M. & Boehlert, C. J. Metal hybrids processed by high-pressure torsion: synthesis, microstructure, mechanical properties and developing trends. *Intl. Mater. Reviews*, DOI: 10.1080/09506608.2021.1922807 (2021).

14. Pouryazdan, M., Kaus, B.J.P., Rack, A, Ershov A. & Hahn, H. Mixing instabilities during shearing of metals. *Nat. Commun*. **8**, 1611 (2017).

15. Sundaram, N. K., Guo, Y. & Chandrasekar, S. Mesoscale Folding, Instability, and Disruption of Laminar Flow in Metal Surfaces. *Phys. Rev. Lett*. **109**, 106001 (2012).

16. Sundaram, N. K., Mahato, A. Guo, Y., Viswanathan, K. & Chandrasekar, S. Folding in metal polycrystals: Microstructural origins and mechanics, *Acta Mater*. **140**, 67-78 (2017).

17. Gola, A., Schwaiger, R., Gumbsch, P. & Pastewka, L. Pattern formation during deformation of metallic nanolaminates. *Phys. Rev. Mater*. **4 (1)**, 013603 (2020).

18. Beckmann, N., et al. Origins of Folding Instabilities on Polycrystalline Metal Surfaces. *Appl. Phys. Rev*. **2 (6)**, 064004 (2014).

19. Beygelzimer, Y. et al. Quantifying Solid-State Mechanical Mixing by High-Pressure Torsion, *J. Alloys Compd*. **878**, 160419 (2021).

20. Beygelzimer, Y., Kulagin, R., Fratzl, P. & Estrin Y. The Earth's Lithosphere Inspires Materials Design. *Adv. Mater*. **33**, 2005473 (2021).

21. Levitas V.I. Phase transformations, fracture, and other structural changes in inelastic materials, *Int. J. Plast*. **140**, 102914 (2021).

22. Fermi, E., Pasta, J., Ulam, S. & Tsingou, M. *Studies of the Nonlinear Problems* (Los Alamos Scientific Lab., N. Mex., 1955).

23. Deutsch, A., Friedl, P., Preziosi, L. & Theraulaz, G. Multi-scale analysis and modelling of collective migration in biological systems. *Phil. Trans. R. Soc. B* **375**, 20190377 (2020).

24. Salcedo-Sanz, S. Modern meta-heuristics based on nonlinear physics processes: A review of models and design procedures, *Phys. Rep*. **655**, 1-70 (2016).

25. Truesdell, C.A. *Rational Thermodynamics* (Springer-Verlag N.Y., 1984).

26. Coxeter, H. S. M. *Introduction to Geometry* (Wiley & Sons, Inc., U.S., 1969).

27. Goetz, A. *Piecewise Isometries — An Emerging Area of Dynamical Systems* (In: Grabner, P. & Woess, W. (eds) Fractals in Graz 2001. Trends in Mathematics, Birkhäuser, Basel, 2003).

28. Yamamoto, K. K., Shearman, T. L., Struckmeyer, E. J. et al. Nature's forms are frilly, flexible, and functional. *Eur. Phys. J. E* **44**, 95 (2021).

29. Juarez, G., Lueptow, R. M., Ottino, J. M., Sturman R. & Wiggins, S. Mixing by cutting and shuffling. *Europhysics Lett*. **91**, 20003 (2010).

30. Sturman, R. The Role of Discontinuities in Mixing. *Adv. Appl. Mech*. **45**, 51-90 (2012).

31. Zaman, Z., Yu, M., Park, P.P. et al. Persistent structures in a three-dimensional dynamical system with flowing and non-flowing regions. Nat Commun 9, 3122 (2018). https://doi.org/10.1038/s41467-018-05508-7

32.     Estrin, Y. & Kubin, L. P. Spatial Coupling and Propagative Plastic Instabilities (In: Continuum Models for Materials with Microstructure, Muehlhaus, H.B. Ed., John Wiley & Sons, N.Y. 1995).

33.     Yu, H. et al. A deformation mechanism of hard metal surrounded by soft metal during roll forming. *Sci. Rep*. **4**, 5017 (2014).

34.     Katz, R., Spiegelman, M. & Holtzman, B. The dynamics of melt and shear localization in partially molten aggregates. *Nature* **442**, 676–679 (2006).

35.     Cui, Y. et al. Plastic Deformation Modes of CuZr/Cu Multilayers. *Sci. Rep*. **6**, 23306 (2016).

36.     Esfahani, S.E., Ghamarian I. & Levitas V.I. Strain-induced multivariant martensitic transformations: A scale-independent simulation of interaction between localized shear bands and microstructure. *Acta Mater*. **196**, 430-443 (2020).

37.     Divinski, S.V., Reglitz, G., Roesner, H., Estrin, Y. & Wilde, G. Ultra-fast diffusion channels in pure Ni severely deformed by equal-channel angular pressing. *Acta Mater*. **59**, 1974-1985 (2011).

38.     Denisov, S., Filippov, A., Klafter J. & Urbakh, M. From deterministic dynamics to kinetic phenomena. *Phys. Rev. E* **69**, 042101 (2004).

39.     Vanossi, A., Bechinger, C. & Urbakh, M. Structural lubricity in soft and hard matter systems. *Nat. Commun.* **11**, 4657 (2020)

40.     Ostermeyer, G.-P. Popov, V. L., Shilko, E. V. & Vasiljeva, O. S. *Multiscale Biomechanics and Tribology of Inorganic and Organic Systems* (Springer, Switzerland, 2021).

41.     Monaghan, J.J. An introduction to SPH. *Comp. Phys. Comm*. **48**, 88-96 (1988).

42.     Radjai, F. & Dubois, F. *Discrete-element modeling of granular materials*.(Wiley-ISTE, London, 2011).

43. Pöschel, T. & Schwager, T. *Computational Granular Dynamics: Models and Algorithms* (Springer, Berlin, 2005).

44. Dmitriev, A.I., Nikonov, A.Y., Filippov, A.E. & Psakhie, S.G. Molecular dynamics study of the evolution of rotational atomic displacements in a crystal subjected to shear deformation. *Phys. Mesomech*. **22**, 375-381 (2019).

45. Filippov A.E. & Gorb, S. *Combined Discrete and Continual Approaches in Biological Modelling* (Springer, Switzerland, 2020).

46. Aifantis E. C. A Concise Review of Gradient Models in Mechanics and Physics. *Front. Phys*. **7**, 239 (2020).

**Supplementary Information**

**Supplementary Information 1. Details of the model used**

The model describes a system of $N$ particles represented by the vector radius $\mathbf{r}_i$, the momentum $\mathbf{p}_i$, and the interaction potential $U(|\mathbf{r}_i - \mathbf{r}_j|)$ corresponding to the following Hamiltonian:

$$H(\mathbf{r}_i, \mathbf{p}_i) = \sum_{i=1}^{N} \mathbf{p}_i^2 / 2m_i + \sum_{i,j=1}^{N} U(|\mathbf{r}_i - \mathbf{r}_j|)/2. \qquad (S1)$$

It is opportune to represent the interaction potential by a pair of the Gauss potentials:

$U(|\mathbf{r}_i - \mathbf{r}_j|) = C_{ij} \exp\{-[(\mathbf{r}_i - \mathbf{r}_j)/c_{ij}]^2\} - D_{ij} \exp\{-[(\mathbf{r}_i - \mathbf{r}_j)/d_{ij}]^2\}$, where $C_{ij}$ and $D_{ij}$ define the magnitude, while $c_{ij}$ and $d_{ij}$ the radii of attraction and repulsion, respectively. The minimisation condition corresponding to an equilibrium reads as follows: $C_{ij} \gg D_{ij}$, $c_{ij} < d_{ij}$. In numerical experiments, the particles are initially placed at random on a rectangle with the side lengths of $[0, L_x]$ and $[0, L_y]$. For the case of shear deformation, it is convenient to employ periodic boundary conditions on the horizontal axis. Accordingly, a particle leaving the interval $[0, L_x]$ is returned to it at the opposite end and the vectors $\mathbf{r}_i - \mathbf{r}_j$ connect particles located within the interval $[0, L_x]$ or the images of escaped particles at the opposite side of the system. On the vertical axis the system is limited by plates at $y = 0$ and $y = L_y$ with reflecting boundary conditions: $U_{up} = C \exp[(y - L_y)/c]$ and $U_{down} = C \exp[-y/c]$. The stronger the inequalities $C \gg C_{ij}$ and $c \ll c_{ij}$, the more rigid and sharp are the walls in relation to other forces and lengths relevant to the system. Since the parameters $C_{ij}$, $D_{ij}$, $c_{ij}$, and $d_{ij}$ form arrays covering all possible combinations, it can be said without loss of generality that the Hamiltonian given by Eq. (S1) describes systems with any number of components, including one-

and two-component systems, which are of interest in this paper. What is required is just to specify the parameters $C_{11}$, $C_{22}$, $C_{12} = C_{21}$, etc.

The simplicity of the equations of motion,

$$m_i \partial \mathbf{v}_i / \partial t = -\partial H(\mathbf{x}_i, \mathbf{p}_i) / \partial \mathbf{p}_i = \mathbf{f}_i, \tag{S2}$$

is deceptive. Summation at every time step over all possible neighbours, including very distant ones and the fictitious image particles stemming from the periodic boundary conditions, is required. In addition, in the case of contact with a thermal bath the boundary conditions need to be replaced with more realistic ones. If the temperature $T$ of the thermal bath at a given boundary (or within the system) is non-zero, an array of random $\delta$-correlated Langevin forces need to be included. These are to satisfy the fluctuation-dissipation theorem in the form of $\langle \xi(t, x_i, y_i) \xi(t', x_j, y_j) \rangle = D \delta(t - t') \delta_{ij}$ $\langle \xi(t, x_i, y_i) \rangle = 0$, where $D = 2\gamma k_B T m / dt$, $k_B$ is the Boltzmann constant, $\gamma$ is a dissipation constant, and $dt$ is the discrete time step. Interacting particles exchange momentum, and hence a dissipation channel acting to equilibrate the relative velocities of particles that happen to be within the distance $c_v$ close to the equilibrium one needs to be introduced. This is done by including an additive force $f_i^v \square \sum_{j=1}^{N} (\mathbf{v}_i - \mathbf{v}_j) \exp\left[-[(\mathbf{r}_i - \mathbf{r}_j)/c_v]^2\right]$ on the *i*-th particle, with yet another dissipation constant $\eta$. As a result, the equations of motion assume the following form:

$$m_i \frac{\partial \mathbf{v}_i}{\partial t} = f_i^r - \eta f_i^v - \gamma \mathbf{v}_i + \xi(t, \mathbf{r}_i) \tag{S3}$$

They can be integrated by using Verlet's method, which conserves the energy of the system at each time step, provided no energy is supplied externally through mechanical work or temperature variation.

**Supplementary Information 2. Calculation of the force**

Here we calculate a measurable quantity: the external force acting on the system. First, we calculate the work produced by the forces acting on each particle of the system or the corresponding power $p_i = \mathbf{f}_i \mathbf{v}_i$. Summation over all particles yields the total power $2VF_{ext}(t) = P(t) = \sum_{i=1}^{N} \mathbf{f}_i \mathbf{v}_i$ expended by a pair of external forces to maintain the motion of the boundary plates with a constant velocity, $V = const$. This yields the total force, $F_{ext}(t)$. Calculating spatially distributed power $p_i = \mathbf{f}_i \mathbf{v}_i$ enables one to evaluate the degree of localisation of excitations and hence the fraction of the regions in which the fragments of the lattice move practically as a whole. As at a particular site the energy is absorbed and released at different points in time, it is useful to characterise these sites by the absolute value of the power, $|p_i| = |\mathbf{f}_i \mathbf{v}_i|$.

For that, we start with plotting a histogram $h(|p_i|)$ for the probability of finding a given value of $|p_i|$. As expected, the function $h(|p_i|)$ has a pronounced maximum at low values of $|p_i|$ associated with small perturbations of the system **almost everywhere within the ordered domains**. The regions of large strain contribute to the total power $2VF_{ext}(t) = P(t) = \sum_{i=1}^{N} \mathbf{f}_i \mathbf{v}_i$ most significantly, but the partial contribution of an individual region is not very significant. As a result, the distribution $h(|p_i|)$ has an extended 'tail' at large $|p_i|$.

For both the locations of the maxima and the asymptotes to be well-resolved at the same a nonlinear function $h(|p_i|)^{1/2}$ is plotted in Fig. S1(a). It can roughly be divided in two regions – left and right of a vertical dashed line that marks a certain characteristic value $p^*$. The fraction of points where $|p_i| > p^*$ holds, i.e., those points where intensive processes occur, evolves with time. The results of this

procedure for a particular realisation of the deformation process are shown exemplary in Fig. S1(b). One can recognise pronounced bursts of activity, which can be associated with the periods of emergence and development of rebuilt structures localised at the edges of domains seen in Supplementary Movie_01.avi. The same periods can also be identified in the time dependence of the total presented in Fig. S1(c). To facilitate the comparison, some of the bursts are marked with dash-dotted lines.

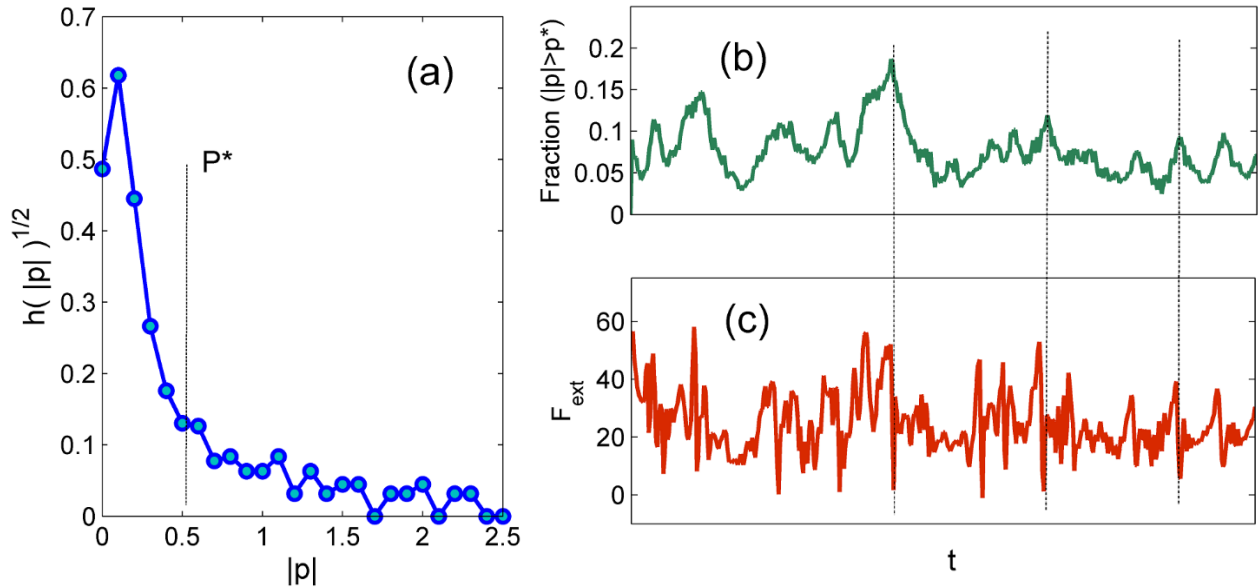

**Figure S1.** Square root $h(|p_i|)^{1/2}$ of the histogram $h(|p_i|)$ (a) and the time dependence of the fraction of the points for which the inequality $|p_i| > p^*$ holds (b). Note the bursts in the magnitude of this fraction corresponding to the peaks in the time dependence of the total force $F_{ext}(t) = P(t)/V$, which is presented in (c).

The function $F_{ext}(t)$ represents the time dependence of the force required to slide a solid on a surface. From this dependence the yield force, or yield stress, can be identified. In our computer experiments, this dependence resembled that for the resistance force for sliding a solid on a surface with random roughness in the stick-slip regime. This is not surprising, as in our case at any given moment at least two layers of the material separated by a rough interface are engaged in a relative frictional motion. In Fig. S2, the occurrence of an upper and a lower yield point can be identified. In metallic materials possessing such a property, propagating localised deformation bands (Lüders bands) occur under tensile loading. A natural question to be asked is why such stick-slip- or Lüders-like effects are not observed or are very weak in real experiments on shear flow of solids. In our opinion, the reason is that shear flow experiments are commonly done on macroscopic objects. In the thermodynamic limit, 'self-averaging' over the individual bursts of the force, which occur in different areas of a macroscopic sample, would suppress pronounced jerky flow.

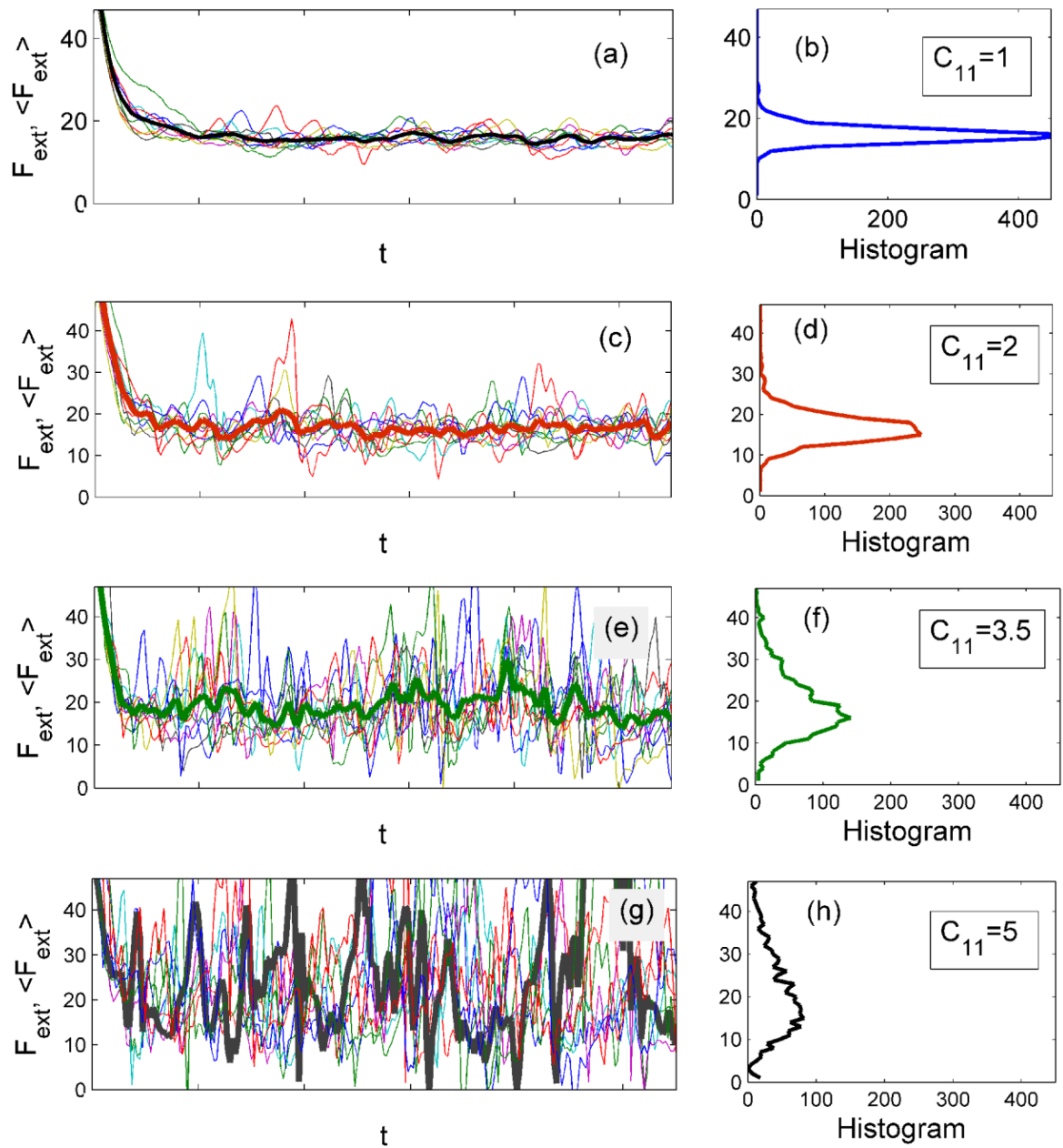

**Figure S2.** The results of averaging of the force over ten realisations of the computer experiments for four different values of the parameter $C_{11}$ representing the stiffness of the interaction potential.

Since one cannot replicate this 'self-averaging' procedure in numerical modelling, it can be replaced with averaging over multiple realisations of the mesoscopic system we are dealing with. Physically, this should correspond to adding up the contributions of different areas of the real body. Fig. S2 displays the results of such averaging over ten realisations for four systems with different values of the stiffness parameter $C_{11}$. The thin lines in the graphs on the left ((a), (c), (e), and (g)) show the time dependence of the force (or the shear, which is equivalent to the force for $V$ = const), whilst the heavy lines represent the results of averaging. The diagrams on the right ((b), (d), (f), and (h)) show the corresponding histograms for the probability of finding an instantaneous value of the force $F_{ext}$ accumulated in all realisations over the entire period of observation.

It should be mentioned that the position of the maximum of a histogram, i.e., the most probable magnitude of the force, is practically stiffness-independent, while due to the tail of a distribution the average force is shifted to larger values when the stiffness is increased. As this shift owes to relatively seldom, but large, maxima of the force, one can conclude that the measured force is governed by sharp discontinuous stick-slip events, which are more pronounced in stiff systems.

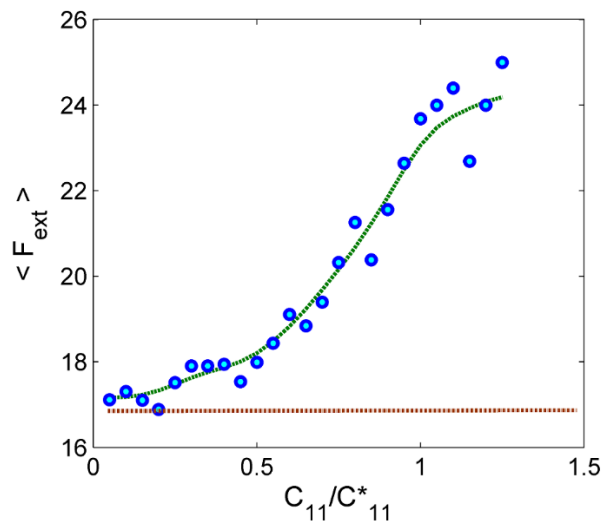

**Figure S3.** Results of computations of the time-averaged force as a function of the stiffness of the system (normalised with respect to a certain value $C_{11}^*$, which was fixed in all numerical experiments). The dash-dotted horizontal line corresponds to a background common to all systems considered. It can be shown that the common background stems from the viscous contribution to the resistance to plastic flow, which in the model considered is present for all systems – whether stiff or soft ('fluid') systems alike. To exclude it, we conducted similar numerical experiments on soft systems, which were otherwise similar to those with a much lower dissipation constant of $\eta = 0.1$ and $\eta = 0.25$. A continual drop of the background level, which is to be subtracted from the force readings when determining the plastic yield limit, is determined for a fixed value of $\eta$, was observed.

In principle, the integral force can be determined both for a large system or, alternatively, through averaging over a large number of realisations. However, analysis of the time dependence of the force, even when averaged over a relatively small number of realisations, shows that it becomes rather flat. For a prolonged steady state, the results of averaging over the histograms in Fig. S2 and the time average converge. This is in keeping with the 'ergodic hypothesis' which states that for a sufficiently large system the average over time and that over the realisations coincide. We used this observation to determine the average force for a representative large number of cases and to establish its dependence on the key parameter, $C_{11}$, which governs the stiffness of the system. The calculated values of the force, which were shown to be in accord with the expected values, are compiled in Fig. S3. The horizontal dash-dotted line corresponding to the background, which is to be subtracted from the data in the case of $\eta = const$.

**Annotations to the videos**

**Supplementary Movie_01.avi**

Typical motion of a solid under shear deformation. The colour code is the same as in Fig. 3. It is seen that initially the solid deforms elastically, the particles' motion smoothly following the displacement of the boundary plates. The corresponding force vs. displacement curve is smooth. When a critical shear is reached, the bonds between neighbouring particles start breaking. However, locally the lattice domains into which the solid gets fragmented move as a whole, switching to a different crystallographic plane only at the contacts between the domains. The profile of the translation velocity averaged over the horizontal axis varies and exhibits different combinations of steps at different points in time. A well-pronounced correlation between the development of cracks and bursts of the power and the total force, as well as waves of localised excitations, is seen on the spatiotemporal velocity diagram.

**Supplementary Movie_02.avi**

The dynamic process leading to the ruptures of a layer of a stiff material sandwiched between two layers of a more compliant material. The particles of the stiff material and the compliant one are marked by filled and open circles, respectively. The same colour code as in Supplementary Movie_01.avi is used. It is seen how the motion of the compliant sub-system in gradually transmitted to the stiff inner layer, engaging it in motion. The diagrams on the right show the distributions of the absolute values of the force and the velocity, averaged along the horizontal axis. It should be noted that the deformation of the inner layer leads to smoothening of the characteristic features of the distributions of the force and the velocity.